\documentclass[aps,pra,twocolumn,preprint,showpacs,preprintnumbers,amsmath,amssymb]{revtex4}

\usepackage{graphicx}
\usepackage{dcolumn}
\usepackage{bm}
\usepackage{color}
\usepackage{slashbox}
\usepackage{amsmath}

\newcommand{\pa}{\partial}

\newcommand{\hanaV}{\mathcal{V}}

\begin{document}

\title{Protection of quantum states from 
disturbance due to random potential by successive translation}
\author{Shumpei Masuda}
\email[]{syunpei710@cmpt.phys.tohoku.ac.jp}
\affiliation{
Department of Physics,
Tohoku University,
Sendai 980, Japan}
\date{\today}
\begin{abstract}
We show a method to protect quantum states from
the disturbance due to 
the random potential by successive rapid manipulations of the quantum states.
The quantum states are kept undisturbed for a longer time than the case of 
the simple trapping with a stationary potential. 
The effective potential, which the quantum states 
feel, becomes uniform when the velocity of the transport is sufficiently large.
It is also shown that the alternating transport of a Bose-Einstein condensate
with the driving potential derived by fast-forward scaling theory
[Masuda and Nakamura, {\it Proc. R. Soc.} A {\bf 466\/}, 1135 (2010)]
can protect it from the disturbance.
\end{abstract}

\maketitle

\section{Introduction}
Technology to control quantum systems is rapidly evolving, and 
various methods to manipulate quantum states have been reported in
Bose Einstein condensates (BEC)\cite{leg,gus,ket,lea},
in quantum computing \cite{nie} and in many
other fields of applied physics. 
For many current and future technologies the acceleration
of controls of quantum systems would be important.
Methods of the acceleration of quantum dynamics and 
quantum adiabatic dynamics or shortcut
to adiabaticity have been proposed, e.g., counterdiabatic protocol
\cite{ric} and frictionless quantum driving \cite{ber3}, invariant-based
inverse engineering \cite{mug1} and fast-forward scaling theory 
\cite{mas1,mas2,mas3}.
These theories make possible to generate target states in short time without
energy excitations at the final time of the manipulations \cite{muga_new}.
Recently applications of these methods 
to the controls of BEC including the transport have been proposed 
theoretically \cite{mug1,mas2,che,mug2,torr,Cam,torr2,mas4}, and been
demonstrated experimentally \cite{Scha,Scha2,Bas}.
The robustness of these protocols have been investigated
\cite{mas1,Che2,torr2}.
In this paper we show 
a novel application and advantage of the acceleration of the 
quantum dynamics.

In actual systems there must be noise
which prevents accurate controls of quantum systems. 
The influence of the random potential to the
static and dynamical properties of the Bose-Einstein 
condensates (BECs) have been studied by using 
the optical speckle potential \cite{Lye,For}.
The protection of quantum states from the influence of the noise and 
disorder would be an important issue  
as well as rapid controls for accurate manipulations of quantum states like 
BECs and for the extension of the range of the quantum controls.
In this paper we use the random potential as the noise
which deforms the wave function in a
trapping potential neglecting the dissipation and 
decoherence due to the environment unlike the 
quantum decoupling of open systems \cite{Vio1,Vio2}.
We show the protection of the quantum states from
the disturbance due to the uncontrollable random potential in the background 
by rapid transport
of the quantum states with the use of the fast-forward scaling theory.
It is shown that the quantum states are kept undisturbed 
for a longer time than the case of 
the simple trapping with a stationary potential because
the effective potential, which the quantum states 
feel, becomes uniform when the velocity of the transport is sufficiently large.
It is numerically exhibited in one dimension 
that the alternating transport of a Bose-Einstein condensate
can protect it from the disturbance.

In Sec.\ref{model} we represent the model. And
the driving potential for the adiabatic transport is reviewed.
In Sec.\ref{Analysis in large velocity limit} we show the
protection of quantum states from the disturbance due to the random potential
in the large-velocity-limit.
In Sec.\ref{Numerical results} it is numerically exhibited that the
fidelity is kept close to unity by the rapid one-way and 
alternative transports.
The protection of a Bose-Einstein condensate is also shown numerically.

\section{Model}
\label{model}
We consider a transport of a particle in one dimension.
The driving potential for the
ideal transport of quantum states without disturbance was derived, 
see e.g. \cite{mas2}.
Suppose that $\Psi_0(x)$ is the wave function of an energy eigenstate 
trapped by the stationary potential $V_0(x)$ 
in the case without the random potential.
The energy is assumed to be zero for the simplicity.
The potential 
\begin{eqnarray}
\hanaV(x,t) = V_0\big{(}x-R(t)\big{)}  -\frac{d^2R(t)}{dt^2} m x.
\label{hanaV}
\end{eqnarray}
can translates the quantum state without energy excitation 
at the final time of the manipulation \cite{mas2}.
The change in $R(t)$ is the displacement of the trapping potential
and the wave function.
The first term in Eq.(\ref{hanaV}) 
corresponds to the translation of the trapping potential.
The second term is the additional potential which is spatially linear.
The wave function of the transported state $\Psi_{FF}$ is represented as
\begin{eqnarray}
\Psi_{FF}(x,t) &=& \Psi_0\big{(}x-R(t)\big{)}\nonumber\\
&&\times\exp\Big{[}i\dot{R}(t)\frac{m}{\hbar}x\Big{]},
\label{psiff}
\end{eqnarray}
where $\dot{R}$ denotes the time-derivative of $R$.
$\dot{R}(t)$ is the velocity of the translation.
$\Psi_{FF}(x,t)$ is a solution of the Schr$\ddot{\mbox{o}}$dinger equation:
\begin{eqnarray}
i\hbar\frac{d\Psi_{FF}}{dt} = 
-\frac{\hbar^2}{2m}\frac{\pa^2\Psi_{FF}}{\pa x^2} + \hanaV(x,t)\Psi_{FF}.
\end{eqnarray}
The additional phase in the wave function in Eq.(\ref{psiff})
vanishes everywhere when the quantum state is stopped and $\dot{R}=0$.

Now we consider the translation under the 
random potential $V_r(x)$.
We assume that the random potential is time-independent.
In general the random potential can cause the disturbance of the wave function.
The total Hamiltonian with the driving potential is represented as
\begin{eqnarray}
H &=& \frac{p^2}{2m} + V_0\big{(}x-R(t)\big{)} \nonumber\\
&&-\frac{d^2R(t)}{dt^2} m x
+ V_r(x).
\label{hami_12_30}
\end{eqnarray}
In the following sections it is shown that 
the disturbance due to the random potential
is restrained by the rapid transport.

\section{Analysis in large velocity limit}
\label{Analysis in large velocity limit}
We show that the fast transport of the quantum states
can reduce the influence of the random 
potential in the case of the constant velocity, $\dot{R}=v$, 
in the large-velocity-limit.
In the analysis 
we use the moving frame which accompanies with the 
trapping potential. 
In the moving frame the third term in Eq.(\ref{hami_12_30}) 
vanishes and the Hamiltonian is represented by
\begin{eqnarray}
H_M = \frac{p^2}{2m} + V_0(x)
+ V_r(x+vt).
\label{HM}
\end{eqnarray}
The trapping potential is stationary in the moving frame while the
random potential is moving with the constant velocity.
We expand the state $\Psi$ 
by the energy eigenstates $|j>$ of the Hamiltonian:
$H_0 = p^2/2m + V_0(x).$
The state is represented as
\begin{eqnarray}
|\Psi(t)> = \sum_j a_j(t)|j>,
\end{eqnarray}
where $H_0|j> = E_j|j>$ and $<j|k> = \delta_{jk}$.
$E_j$ is the energy of the $j$th state.
$|j>$ is time-independent.
The Schr$\ddot{\mbox{o}}$dinger equation leads to
the equations of the coefficients:
\begin{eqnarray}
\pa_t a_j(t) = -\frac{i}{\hbar} E_j a_j(t)
-\frac{i}{\hbar}\sum_{k}V_{jk}(t)a_k(t),
\end{eqnarray}
where $V_{jk}(t)$ is defined by
\begin{eqnarray}
V_{jk}(t)&\equiv& <j|V_r(x+vt)|k>\nonumber\\
&=&\int_{-\infty}^\infty\phi_j^\ast(x)\phi_k(x)
V_r(x+vt)dx,
\end{eqnarray}
with $\phi_j(x)=<x|j>$.
We suppose that the initial state is the $n$th energy eigenstate $|n>$, that is,
$a_j(0) = \delta_{jn}$.
$a_j(t)$ satisfies the integral equation:
\begin{eqnarray}
a_j(t) &=& e^{-\frac{i}{\hbar}E_j t}\delta_{jn}\nonumber\\
&&-\frac{i}{\hbar}\sum_k\int_{0}^t ds\ e^{-\frac{i}{\hbar}E_j(t-s)} V_{jk}(s)a_k(s).
\nonumber\\
\label{eq1_13_1}
\end{eqnarray}
For $m\ne n$ we obtain the 1st order approximation of $a_m(t)$
by substituting $j=m$ and the $0$th order solution 
$a_k(s) = \delta_{kn}\exp[-iE_ks/\hbar]$
in Eq.(\ref{eq1_13_1}) as
\begin{eqnarray}
a_m(t) &\simeq& 
-\frac{i}{\hbar}\sum_k\int_{0}^t ds\ e^{-\frac{i}{\hbar}E_m(t-s)}\nonumber\\
&&\times V_{mk}(s)e^{-\frac{i}{\hbar}E_k s}\delta_{kn}\nonumber\\
&=& -\frac{i}{\hbar}e^{-\frac{i}{\hbar}E_mt}\nonumber\\
&&\times\int_{0}^t ds\ e^{\frac{i}{\hbar}(E_m-E_n)s}
V_{mn}(s).
\label{eq1_24_1}
\end{eqnarray}
We assume that a finite number of the energy eigenstates can 
have the dominant contribution to the transition from the $n$th eigenstate
in the 1st order approximation
and the others are negligible because of the sufficiently small $V_{mn}$.
Hereafter we focus on the energy eigenstates which may have the dominant 
contribution.
We divide the time integral in Eq.(\ref{eq1_24_1}) into the intervals as 
\begin{eqnarray}
\int_0^t ds = \int_0^{\Delta t}ds + \int_{\Delta t}^{2\Delta t}ds +
\cdots + \int_{t-\Delta t}^tds.\nonumber\\
\end{eqnarray}
We take $\Delta t$ short enough so that 
$\exp[i(E_m-E_n)s/\hbar]$ can be regarded as constant
in the interval for any $m$, that is, 
\begin{eqnarray}
\omega_{mn}\Delta t \ll 1,
\label{con1}
\end{eqnarray}
where $\omega_{mn} = (E_m-E_n)/\hbar$. 
For example the $q$th integral with respect to $s$ 
of the last line in Eq.(\ref{eq1_24_1}) is represented as
\begin{eqnarray}
&&e^{\frac{i}{\hbar}\omega_{mn} q\Delta t}
\int_{(q-1)\Delta t}^{q\Delta t}dsV_{mn}(s)\nonumber\\
&&=e^{\frac{i}{\hbar}\omega_{mn} q\Delta t}
\int_{(q-1)\Delta t}^{q\Delta t}ds\int_{-\infty}^{\infty}dx\nonumber\\
&&\ \ \times\phi_m^\ast(x)\phi_n(x)
V_r(x+vs)
\end{eqnarray}
The integration with respect to $s$ is rewritten with $\tau\equiv x+vs$ as
\begin{eqnarray}
&&\int_{(q-1)\Delta t}^{q\Delta t}ds V_r(x+vs) \nonumber\\
&&= \Delta t\frac{1}{v\Delta t}
\int_{x+v(q-1)\Delta t}^{x+vq\Delta t}V_r(\tau)d\tau.
\end{eqnarray}
We define the effective potential $\bar{V}_r(x,q)$ by
\begin{eqnarray}
\bar{V}_r(x,q) = \frac{1}{v\Delta t}
\int_{x+v(q-1)\Delta t}^{x+vq\Delta t}V_r(\tau)d\tau,
\label{Vbar}
\end{eqnarray}
which is the average of the random potential in the interval: 
$\Delta x = v\Delta t$.
Suppose that $l$ is the length of $v\Delta t$ with which $\bar{V}_r(x)$
can be regarded as uniform in the region that wave function is localized.
Thus if
\begin{eqnarray}
v\Delta t > l,
\label{con2}
\end{eqnarray}
we have
\begin{eqnarray}
&&a_m(t)\simeq -\frac{i\Delta t}{\hbar}
e^{-\frac{i}{\hbar}E_m t}\sum_qe^{i\omega_{mn} q\Delta t}\nonumber\\
&&\times\int_{-\infty}^\infty dx\
\phi_m^\ast(x)\phi_n(x)\bar{V}_r(x,q) \simeq 0,
\end{eqnarray}
due to the orthogonality of $\phi_n$ and $\phi_m$.
Therefore in the large $v$ limit, we see no transition among the 
energy eigenstates.
From the conditions in Eqs.(\ref{con1}) and (\ref{con2})
We obtain a criterion of $v$:
\begin{eqnarray}
v \gg \omega_{mn} l
\end{eqnarray}
with which the level transitions due to the 
random potential do not occur.

\section{Numerical results}
\label{Numerical results}
We numerically exhibit the protection of quantum states 
by the one-way translation with the 
constant velocity and the alternating translation in one dimension.
\subsection{Translation with constant velocity}
\label{Translation with constant velocity}
\begin{figure}[h!]
\begin{center}
\includegraphics[width=8cm]{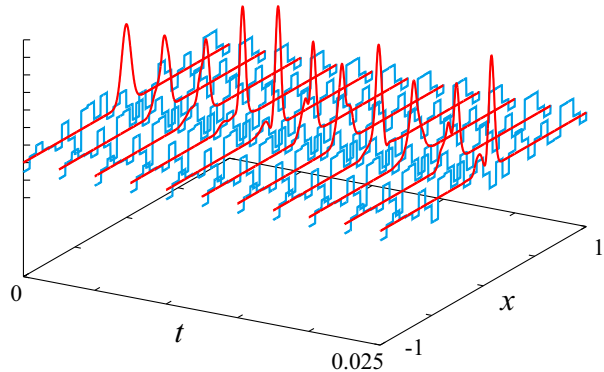}
\end{center}
\caption{\label{fig:epsart} Snapshots of the time-evolution of 
$|\Psi|^2$ under random potential for $\dot{R}=0$.
Red and blue lines correspond to $|\Psi|^2$ and the random potential,
respectively. The parameters are taken as $m=1$, $\hbar=1$, $\omega=256$,
$\eta=0.065$ and $W=1000$.}
\label{fid_p2}
\end{figure}

We numerically simulate the transport with the constant velocity 
under the random potential.
The fidelity of the quantum state is calculated during the time-evolution.
We chose the trapping potential $V_0$ in Eq.(\ref{hami_12_30}) 
as the harmonic potential:
\begin{eqnarray}
V_0(x) = \frac{m\omega^2}{2}x^2.
\label{harm}
\end{eqnarray}
The initial state is taken as the ground state in the harmonic potential 
multiplied by a phase factor as
\begin{eqnarray}
\Psi(x,0) = \Big{(}\frac{m\omega}{\pi\hbar}\Big{)}^{\frac{1}{4}}
\exp\Big{[}{-\frac{m\omega}{2\hbar}x^2+i\dot{R}\frac{m}{\hbar}x}\Big{]},
\nonumber\\
\label{inip}
\end{eqnarray}
where $\dot{R}$ is the velocity of the transport and is constant.
$\Psi(x,t)$ is deformed from the exactly transported state $\Psi_{FF}(x,t)$
in Eq.(\ref{psiff}) due to the random potential, while they
coincide with each other at the initial time.
The random potential has the rectangular form with the width $\eta$
as shown in Fig.\ref{fid_p2}.
The random potential takes the value from $-W/2$ to $W/2$.
We drive the wave function in $x$-direction by moving the 
harmonic trap with the constant velocity (see Eq.(\ref{hami_12_30})).
In the numerical simulation 
we use the moving frame which accompanies with the trapping potential.
The corresponding Hamiltonian is given by Eq.(\ref{HM}).

\begin{figure}[h]
\begin{center}
\includegraphics[width=7cm]{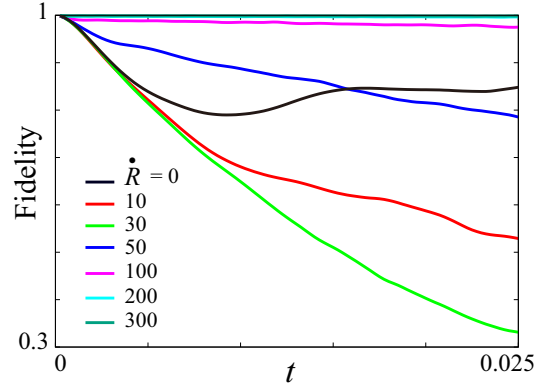}
\end{center}
\caption{\label{fig:epsart} Time dependence of fidelity in one-way driving. 
The fidelity is averaged over the dynamics with 100 different random potentials.
$\dot{R}$ is the velocity of the centre of trapping potential.
Other parameters are the same as Fig.\ref{fid_p2}.}
\label{fid_com}
\end{figure}

The snapshots of the time-evolution of 
$|\Psi|^2$ under random potential are shown in Fig.\ref{fid_p2} for $\dot{R}=0$
which corresponds to the simple trapping with the stationary potential.
The harmonic potential is not shown in the figure.
The parameters are taken as $m=1$, $\hbar=1$, $\omega=256$,
$\eta=0.065$ and $W=1000$.
The wave function is deformed due to the random potential.
The time dependence of the 
fidelity is shown in Fig.\ref{fid_com} for various values of $\dot{R}$.
The fidelity is defined by $|<\Psi_{FF}(t)|\Psi(t)>|$
where $|\Psi_{FF}>$ in Eq.(\ref{psiff}) is the exactly transported state 
without the disturbance of the random potential.  
The fidelity is averaged over the dynamics with 100 different random potentials.
We see the decrease of the 
fidelity with time due to the disturbance by the random potential
for $\dot{R}=0$.
The fidelity for $\dot{R}=10$ and $30$ is
lower than that of $\dot{R}=0$.
This is because that the relative time-dependence of the random potential 
enhances the energy transition. 
However, for the sufficiently large velocity, 
$\dot{R}\ge 100$, the decrease of the fidelity is 
apparently restrained compared to the case of $\dot{R}=0$.
The continuous transport reduces the influence of the random potential as 
theoretically predicted in the previous section.
The variance of the fidelity is about 0.01 for $\dot{R}=100$.
The variance tends to be smaller for larger $\dot{R}$.

\begin{figure}[h]
\begin{center}
\includegraphics[width=8cm]{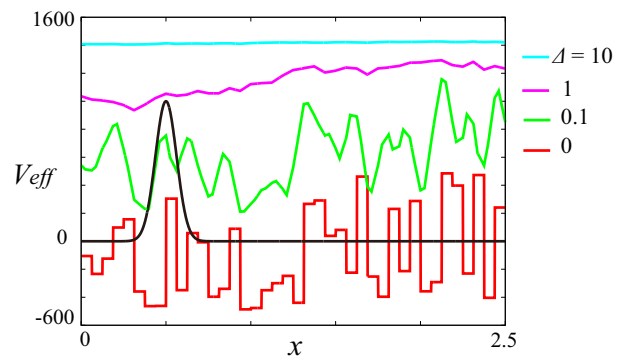}
\end{center}
\caption{\label{fig:epsart} 
Effective potential $V_{eff}(x, \delta l)$ for $\delta l=0, 0.1, 1, 10$. 
$V_{eff}$ for $\delta l > 0$ are shifted upward for the comparison.
Other parameters are the same as Fig.\ref{fid_p2}.}
\label{Veff}
\end{figure}

$\bar{V}_r$ in Eq.(\ref{Vbar}) is regarded as the effective potential that the 
quantum state feels. 
To show the property of the effective potential we calculate 
\begin{eqnarray}
V_{eff}(x,\delta l)\equiv  \frac{1}{\delta l}
\int_{x}^{x+\delta l}V_r(\tau)d\tau,
\label{Veff1}
\end{eqnarray}
which corresponds to $\bar{V}_r$ for $q=0$.
The distance $\delta l$ corresponds to $v\Delta t$.
The effective potential $V_{eff}(x, \delta l)$ for the various values of 
$\delta l$ are shown in Fig.\ref{Veff}.
For large $\delta l$ ($\delta l=1$ and $10$) 
the effective potential become smooth and uniform compared to the original
random potential.
This property of the effective potential explains the
decrease of the influence of the random potential. 
The distance $\delta l=1$ corresponds to $\dot{R}=256$
if we relate the distance $\delta l$, the time-interval $\Delta t$
and the frequency $\omega$ by
$\delta l=\dot{R} \Delta t$ and $\Delta t = 1/\omega$.

\subsection{Alternating translation}
\label{Alternating translation}
In actual systems it is impossible to keep translating the quantum state
in one direction.
Here we show that the alternating translation also
protects quantum states from the disturbance due to the random potential 
in one dimension.
The trapping 
potential $V_0(x)$ is the harmonic potential as Eq.(\ref{harm}),
and the initial state is the ground state given by Eq.(\ref{inip})
with $\dot{R}=0$.
We choose the time dependence of $R(t)$ as  
\begin{eqnarray}
R(t) = \frac{L}{2}\{1-\cos(\omega_R t)\}. 
\label{Rt}
\end{eqnarray}
We continuously translate the quantum state back and forth
by translating the trapping potential
and simultaneously tuning the spatially linear potential as Eq.(\ref{hanaV}).
The center of the wave packet is oscillated between $x=0$ and $x=L$ 
periodically.
The random potential is the same as 
Sec.\ref{Translation with constant velocity}. 
In the numerical simulation 
we use the moving frame which accompanies with the original 
trapping potential $V_0$.

\begin{figure}[h]
\begin{center}
\includegraphics[width=8cm]{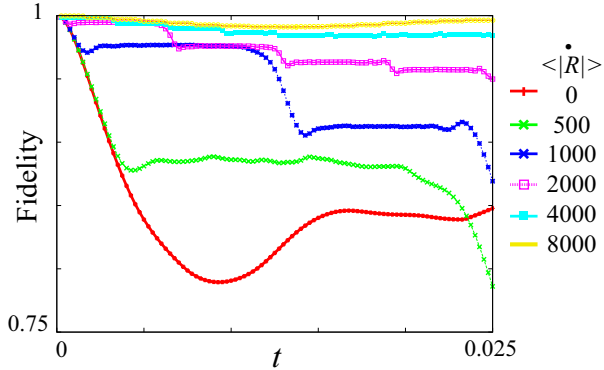}
\end{center}
\caption{\label{fig:epsart} Time dependence of fidelity in the
alternating translation for various values of $<|\dot{R}|>$. 
The fidelity is averaged over the dynamics with 100 different random potentials.
Other parameters are the same as Fig.\ref{fid_p2}.}
\label{fid_com2}
\end{figure}

The time dependence of the fidelity is shown in Fig.\ref{fid_com2} for various 
values of $\omega_R$.
The corresponding value of the 
time-average of $|\dot{R}|$ denoted by $<|\dot{R}|>$ are shown in the figure. 
We put $L=2$. 
Other parameters are the same as Fig.\ref{fid_p2}.
The fidelity is averaged over the dynamics with 100 different random potentials.
The variance of the fidelity is about 0.015 for $<|\dot{R}|>=4000$.
The variance tends to be smaller for larger $<|\dot{R}|>$.
It is seen that the alternating translation with 
the sufficiently large frequency 
can reduce the influence of random potential.
The rapid falls of the fidelity occur at the 
time when the trapping potential is turned and starts to move in the 
opposite direction.
Since the velocity becomes small at such time, 
the quantum state is affected by the random potential and the fidelity 
decreases.

\subsection{Bose-Einstein condensates}
\begin{figure}[h]
\begin{center}
\includegraphics[width=8cm]{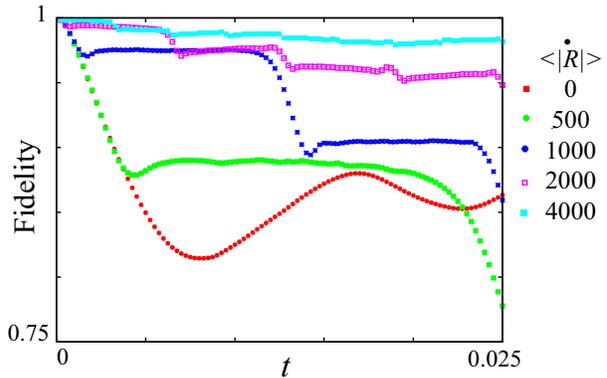}
\end{center}
\caption{\label{fig:epsart} Time dependence of fidelity in 
alternating translation for $c=10$. 
Other parameters are the same as Fig.\ref{fid_com2}.
}
\label{fid_com_non}
\end{figure}

Here we apply the method to reduce the influence of the random potential to
Bose-Einstein condensates (BECs).
We assume that the system is governed by the
Gross-Pitaevskii (GP) equation:
\begin{eqnarray}
i\hbar\frac{d\Psi}{dt} = 
-\frac{\hbar^2}{2m}\frac{\pa^2\Psi}{\pa x^2} + \hanaV(x,t)\Psi
+ c|\Psi|^2\Psi,\nonumber\\
\end{eqnarray}
where $\Psi(x,t)$ is the macroscopic wave function,
and $c$ is the coupling parameter.
It has been shown that the BECs are also transported 
without energy excitation by the same driving 
potential in Eq.(\ref{hanaV}) \cite{mas2,torr3,mas4}.
Let us suppose that 
a BEC wave packet is trapped by the harmonic potential subjected
to the random potential which is the same as the previous sections.
The initial state is the ground state obtained numerically 
in the stationary harmonic potential without the random potential.
We alternately translate the wave packet by using the driving potential.
The time dependence of $R(t)$ and the parameters are same 
as Sec.\ref{Alternating translation}.
In Fig.\ref{fid_com_non} the time dependence of the fidelity 
is shown for $c=10$. 
The fidelity shows the similar time-dependence to the case with $c=0$.
The results show 
that the disturbance of BECs can be suppressed by the repeated driving
of the wave packet.

\section{summary and discussion}
We have presented the protection of quantum states from the disturbance due to
the random potential by the continuous transport with the use of
the driving potential derived by the theory
of the acceleration of adiabatic dynamics.
We analytically showed that the fast translation of the 
quantum states can suppress 
the influence of the random potential because
the effective potential which the quantum states feel becomes uniform when 
the velocity of the transport is sufficiently large.
We emphasize that the velocity of the transport does not have to be fast 
enough so that the effective potential vanishes.

We have 
numerically exhibited the protection of the quantum states 
by the continuous translation with the 
constant velocity and the alternating translation of
the the ground state in the harmonic potential.
The decrease of the fidelity is clearly
restrained by the fast-driving compared to 
the simple trapping with a stationary potential, while 
the protection effect is not monotonously increased with 
the velocity in the small-velocity-region. 
It has been shown that the same technique is effective also for the protection
of BECs from the influence of the random potential.

The optical speckle potential
were used to investigate the properties of 
BECs under the random potential experimentally \cite{Lye,For}.
It is expected that such system would be useful to 
investigate the present method experimentally because 
the strength of the random potential 
can be controlled by tuning the intensity of the laser. 
In this paper we did not consider the dissipation and decoherence effects.
Applications of the present method in the system with
the dissipation and decoherence would be studied in the future also 
with the effect of the time-dependence of the random potential. 
We assumed that the driving potential is controlled without error, while
for rapid manipulation the control of the potential itself can cause 
additional noises in actual systems.
The investigation of the robustness of the present technique is a future issue.

\begin{acknowledgments}
The author thanks K. Nakamura for useful discussions and comments.
The author thanks global COE program 
``Weaving Science Web beyond Particle-Matter Hierarchy''
for its financial support.
The author is 
also financially supported by Grants-in-Aid for Centric Research of 
Japan Society for Promotion of Science.
\end{acknowledgments}

\end{document}